# Exploring Strategies for Classification of External Stimuli Using Statistical Features of the Plant Electrical Response

Shre Kumar Chatterjee[a], Saptarshi Das[a,*], Koushik Maharatna[a], Elisa Masi[b], Luisa Santopolo[b], Stefano Mancuso[b] and Andrea Vitaletti[c,d]

a) School of Electronics and Computer Science, University of Southampton, Southampton SO17 1BJ, United Kingdom
b) Department of Agri-food Production and Environmental Science (DISPAA), University of Florence, viale delle Idee 30, 50019 Sesto Fiorentino, FIorence, Italy
c) WLAB S.r.L., via Adolfo Ravà 124, 00142 Rome, Italy
d) DIAG, SAPIENZA Università di Roma, via Ariosto 25, 00185 Rome, Italy

**Author's Emails:**

skc105@ecs.soton.ac.uk (S.K. Chatterjee)

sd2a11@ecs.soton.ac.uk, s.das@soton.ac.uk (S. Das*)

km3@ecs.soton.ac.uk (K. Maharatna)

elisa.masi@unifi.it (E. Masi)

luisa.santopolo@unifi.it (L. Santopolo)

stefano.mancuso@unifi.it (S. Mancuso)

andrea.vitaletti@w-lab.it (A. Vitaletti)

**Abstract:**

Plants sense their environment by producing electrical signals which in essence represent changes in underlying physiological processes. These electrical signals, when monitored, show both stochastic and deterministic dynamics. In this paper, we compute 11 statistical features from the raw non-stationary plant electrical signal time series to classify the stimulus applied (causing the electrical signal). By using different discriminant analysis based classification techniques, we successfully establish that there is enough information in the raw electrical signal to classify the stimuli. In the process, we also propose two standard features which consistently give good classification results for three types of stimuli - Sodium Chloride (NaCl), Sulphuric Acid ($H_2SO_4$) and Ozone ($O_3$). This may facilitate reduction in the complexity involved in computing all the features for online classification of similar external stimuli in future.

## 1. Introduction

Plants produce electrical signals, when subjected to various environmental stimuli [1–7]. These electrical signals in essence represent changes in underlying physiological processes influenced by the external stimuli. Thus, analysing such plant electrical signals may uncover possible signatures of the external stimuli embedded within the signal. The stimuli may vary from different light conditions, burning, cutting, wounding, gas or liquid [8] etc. This opens up the possibility to use such analysis techniques to turn a green plant into a multiple-stimuli sensing biological sensor device [9]. If such an association between the external stimuli and the resulting plant electrical signal could be made, then it may serve the purpose of holistic monitoring of environmental constituents at a much cheaper cost (because of abundance of plants) thereby eliminating the need to install multiple individual sensors to monitor the same external stimuli. In this work, we attempt to explore the possibility of classifying three external stimuli - Sodium Chloride (NaCl), Sulphuric Acid ($H_2SO_4$) and Ozone ($O_3$), from the electrical signal response of plants as the first step towards that goal. Here, we chose heterogeneous stimuli that reproduce some of the possible environmental pollutants e.g. $H_2SO_4$ is a major component of acid rain. Ozone is a tropospheric air pollutant and is the main component of smog. Salinization often results from irrigation management practices or treatment of roads with salt as de-icing agent and can be linked to environmental soil pollution. These three stimuli – NaCl, $H_2SO_4$, $O_3$ are specifically chosen to study the change in plant physiological response to represent the effect of environmental pollution.

Electrical signals were collected from a number of tomato (*Solanum lycopersicum*) and cucumber (*Cucumis sativus*) plants using NaCl, $H_2SO_4$ and $O_3$ as stimulus in controlled settings. Multiple experiments were conducted for each stimulus to ensure the repeatability of the electrical signal response each time. We then extracted 11 statistical features from these plant signal time series in order to investigate the possibility of accurate detection of external stimulus through a combination of these features and simple discriminant analysis classifiers. We believe this work will not only form the backbone of using plants as environmental biosensors [9], [10] but also open up a new field of further exploration in plant signal behaviours with meaningful feature extraction and classification similar to the studies done using other human body electrical responses like Electrocardiogram (ECG), Electroencephalogram (EEG) and Electromyogram (EMG) [11].

Although there have been few recent attempts on signal processing, feature extraction and statistical analysis using plant electrical responses [12–18], there has been no attempt to associate features extracted from plant electrical signals to different external stimuli. The focus of our work is to address this gap. Here, we analysed the statistical behaviour of raw electrical signals from plants similar to previous studies on raw non-stationary biological signals which exhibit random fluctuations such as EMG/EEG, adopting a similar approach to develop a classification system [19–22]. The present paper reports the first exploration of its kind, aiming at finding meaningful statistical feature(s) from segmented plant electrical signals which may contain some signature of the stimulus hidden in them, in different extents.

As a first exploration, this work focussed on the ability to classify the stimuli by only looking at a small segment of raw plant electrical response. The questions which arise in order to explore this possibility of classification are: 1) which features give a good discrimination between the stimuli, and 2) which type of simple classifier will give a consistently good result. The simplicity of the classifier is important issue here because our ambition is to run it on resource constrained embedded systems, such as sensor nodes in future. In order to tackle the first question, we start by using 11 statistical features which have been used in other biological signals as well (e.g. EEG, ECG and EMG) [11]. We here explore which feature alone (univariate analysis) or feature combinations (bivariate analysis) consistently indicate towards that particular signature of the stimulus. In order to answer the second question, we will start with a simple discriminant analysis classifier and then its other variants to observe the average classification rate.

## 2. Material and methods
### (a) Stimulus and experimental details

Here we try to develop a classification strategy to detect three different stimuli *viz.* $O_3$, $H_2SO_4$, NaCl. Four set of experiments were conducted with $H_2SO_4$, NaCl 5ml and 10ml each as stimuli, as shown in Table 1. For each stimuli mentioned above, a between subjects design for experiments were setup where four different tomato plants (similar age, growing conditions and heights) were used with each plant being exposed to the stimulus only once. Thus for 12 experiments, 12 tomato plants were used. For Ozone as stimuli, six cucumber plants and two tomato plants were used for eight experiments with each plant being subjected to only one experiment but multiple application of the stimulus.

Table 1: Different stimulus, plants species and number of data-points (each capturing 11 statistical measures of 1000 samples) used for the present study

| **Stimulus** | **Plant species used** | **Concentration and application** | **Number of data-points** |
|---|---|---|---|
| Ozone ($O_3$) | Tomato/Cucumber | 16 ppm for a minute, every 2 hours | 1881 |
| Sulphuric acid ($H_2SO_4$) | Tomato | 5 ml of $H_2SO_4$ 0.05M in the soil once | 496 |
| Sodium Chloride (NaCl) – 5ml | Tomato | 5 ml of NaCl 3M in the soil once | 812 |
| Sodium Chloride (NaCl) – 10ml | Tomato | 10 ml of NaCl 3M in the soil once | 612 |

For each plant, we used three stainless steel needle electrodes - one at the base (reference for background noise subtraction), one in the middle and the other on top of the stem as shown in Figure 1. The electrodes were 0.35 mm in diameter and 15 mm in length, similar to those used in EMG from Bionen S.A.S. and were inserted around 5 – 7 mm into the plant stem so that the sensitive active part of the electrodes (2mm) are in contact with the plant cells [8]. The electrodes were connected to the amplifier-Data Acquisition (DAQ) system in a same way previously studied in [8]. Plants were then enclosed in a plastic transparent box with proper openings to allow the presence of cables and inlet/outlet tubes, and exposed to artificial light conditions (LED lights responding to plant's photosynthetic needs, mimicking a day/night cycle of 12 hours). Each experiment was conducted in a dark room to avoid external light interferences. The whole setup was then placed inside a Faraday cage to limit the effect of electromagnetic interference as shown in Figure 1.

After the insertion of the electrodes into the plant, we waited for about 45 minutes to allow the plant(s) to recover before starting the stimulations. Electrical signals acquired by the electrodes were provided as input to a 2-channel high impedance ($10^{15}$ Ω) electrometer (DUO 773, WPI, USA) while data recording was carried out through 4-Channel DAQ (LabTrax, WPI) and its dedicated software LabScribe (WPI) [23]. The sampling frequency was set as 10 samples per second for all the recordings. For the treatments with liquid, sulphuric acid (5 ml $H_2SO_4$, 0.05M) or sodium chloride (5 or 10 ml NaCl 3M), a syringe placed outside of the Faraday cage and connected to a silicone tube inserted into the plant soil, was used to inject the solution as shown in Figure 2(a). $O_3$, produced by a commercial ozone generator (mod. STERIL, OZONIS, Italy), [24] was injected into the box through a silicone tube (1 minute spray every 2 hours, 16 ppm), while a second outlet tube threw the Ozone from the box to the chemical hood as shown in Figure 2(b). The concentration of Ozone inside the box was monitored using a suitable sensor.

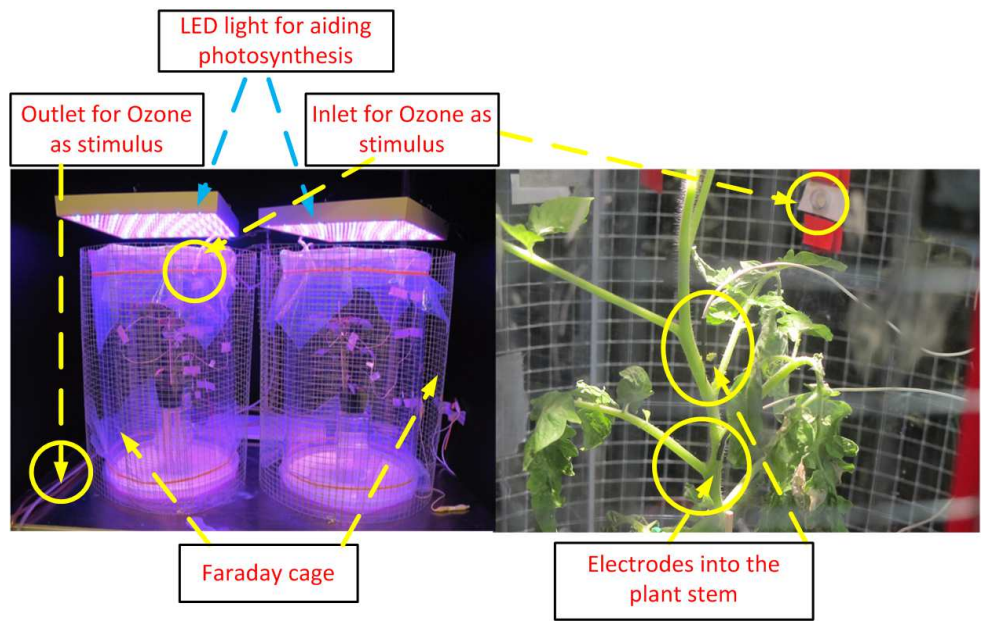

Figure 1: Experimental setup showing a tomato plant inside a plastic transparent box, kept inside a Faraday cage. The placement of the electrodes on the stem is also shown.

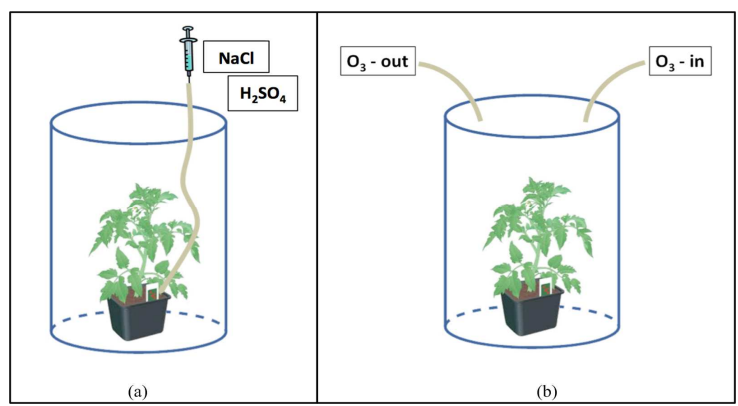

Figure 2: Tubes system for introducing pollutants inside the box. (a) For the treatments with $H_2SO_4$ or NaCl, a syringe placed outside of the Faraday cage and connected to a silicone tube inserted into the plant soil was used to inject the solution at various concentrations. (b) Ozone was injected into the box through a silicone tube, while a second outlet tube threw the ozone from the box to the chemical hood.

*(b) Data processing and segmentation*

Each dataset was obtained after one (H$_2$SO$_4$, NaCl 5 ml and 10 ml) or multiple (O$_3$) application of that particular stimulus. This is illustrated in Figure 3 where the application of stimulus is marked by a vertical dotted line with the post stimulus part of the time series on the right side and the background or pre-stimulus part indicated on the left side of the line. In the case of O$_3$, multiple application of the stimulus is shown by multiple markers.

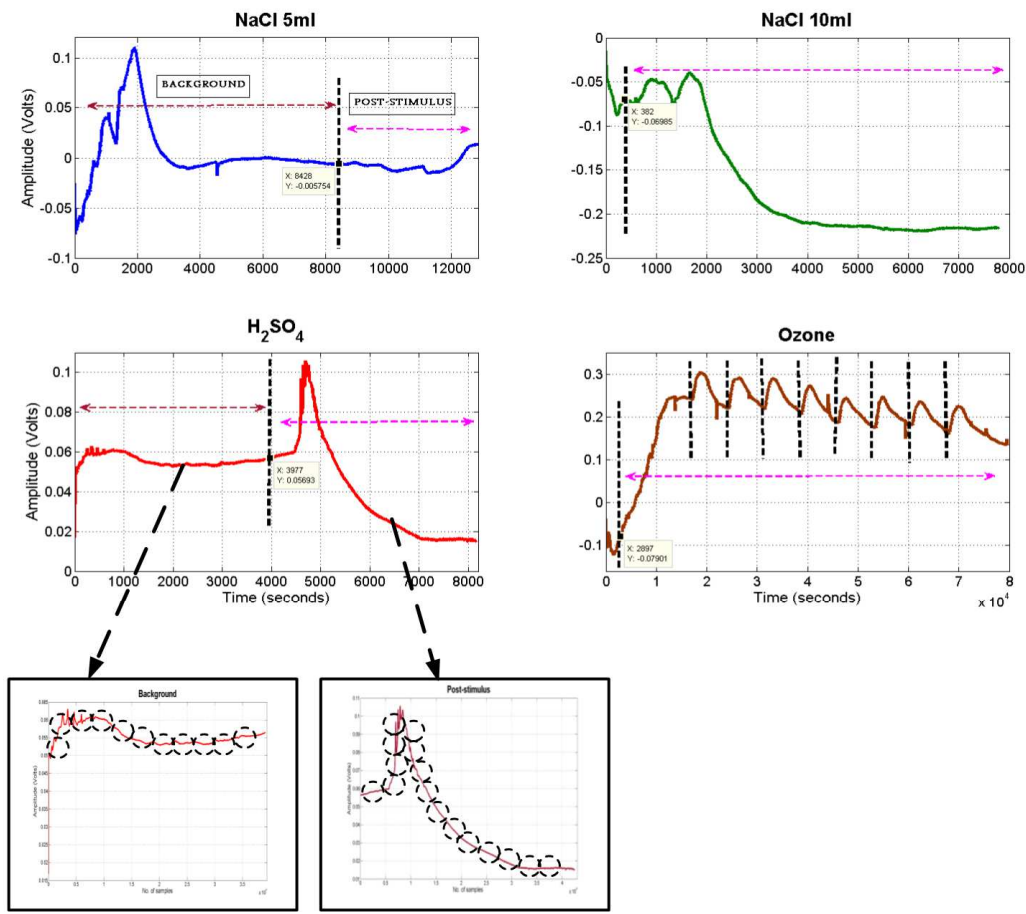

Figure 3: (Top) The vertical dotted lines mark the application time of the four stimuli. (Bottom) Separating the plant electrical signal into background and post-stimulus parts and then dividing them into smaller blocks of 1000 samples, as shown by dashed circles).

As a general observation, from Figure 3 we can see that there are sudden spiking changes in the signal after the application of H$_2$SO$_4$ and Ozone as stimulus. However for the NaCl 5ml and 10 ml stimuli, the changes in the electrical signal response are relatively slow. Thereafter, for each experiment, we divided the data such that we have a post stimulus part of the signal as well as the background (pre-stimulus) part. In case of O$_3$ where multiple stimuli were applied, we divided the data such that the signal duration between consecutive applications of the stimuli is a separate post-stimulus response. This way, we ended up having several post and pre-stimuli datasets for all the four stimuli. Next, each of these datasets was segmented into blocks of fixed window length of 1000 samples (100 seconds) which is shown in Figure 3.

The reason for this data segmentation is to facilitate batch processing of large volumes of data acquired during continuous monitoring. We extracted 11 statistical features from these small chunks of 1000 samples and wanted to explore if the features from such small chunks give enough information to the classifier to discriminate which stimulus that particular data chunk (time period) belonged to. Again, a successful classification of the stimulus from the features of such a small signal block will enable a fast decision time. This is due to smaller buffer-size for batch processing compared to the whole length of the signal acquisition thereby making it easier for possible online implementation in future. Since this is the first exploration of its kind, we stick to 1000 samples only, for extracting statistical features which give sufficiently good classification accuracy but there is scope for further exploration of an optimum window length to classify the stimulus. The classifier was trained using only the blocks of samples belonging to the post-stimulus part of the plant signal. The pre-stimulus part was also divided into similar segments in order to study the effect of the background for different plants under different experimental condition.

The stimulus induced plant signals have both deterministic and random dynamics i.e. local and global variations in amplitudes and different statistical measures of smaller data segments [6], [9], [10], [25], [26]. The research question which we try to answer through the present exploration is – is it possible to identify the stimulus by only looking at the statistical behaviour of small segments of the plant electrical response? A successful answer to this question would pave the way of conceptualising an electronic sensor module in future for classifying the environmental stimulus. This sensor module can be fitted on the plant for batch processing of segmented plant signals, statistical feature extraction and classification, without much memory requirement in future applications.

### *(c) Statistical feature extraction from segmented time series*

Here, we started with 11 features which are predominantly used in the analysis of other biological signals [27]. Different descriptive statistical features like mean ($\mu$), variance ($\sigma^2$), skewness ($\gamma$), kurtosis ($\beta$) as given in (1) and Interquartile range ($IQR = Q_3 - Q_1$, i.e. the difference between the 1st and 3rd Quartile) were calculated.

$$\mu = E[x_i], \sigma^2 = E[x_i - \mu]^2, \gamma = E\left[(x_i - \mu)/\sigma\right]^3, \beta = E\left[(x_i - \mu)/\sigma\right]^4 \tag{1}$$

In the calculation of four basic moments in (1), $x_i$ is the segmented raw electrical signals each of them containing 1000 samples and *E[.]* is the mathematical expectation operator. Apart from these five, the remaining six features taken are – Hjorth mobility, Hjorth complexity, detrended fluctuation analysis (DFA), Hurst exponent, wavelet packet entropy and average spectral power which are briefly described below.

### *Hjorth's parameters*

The Hjorth mobility and complexity, described in [28], quantify a signal from its mean slope and curvature by using the variances of the deflection of the curve and the variances of their first and second derivatives. Let the signal amplitudes at discrete time instants be $a_n$ at time $t_n$. The measures of the complexity of the signal is based on the second moments in time domain of the signal and the signal's first and second derivatives. The finite differences of the signal or time derivatives can be viewed in (2).

$$d_n = a'_n = a_{n+1} - a_n, \text{ where } n = 1, 2, \cdots, (N-1) \text{ and}$$

$$d'_n = a''_n = a'_{n+1} - a'_n, \text{ where } n = 1, 2, \cdots, (N-2) \tag{2}$$

The variances are then computed as (3) [29].

$$\sigma_a^2 = \frac{1}{N}\sum_{n=1}^{N} a_n^2, \; \sigma_d^2 = \frac{1}{(N-1)}\sum_{n=1}^{N-1}(d_n - d_{n-1})^2, \; \sigma_{dd}^2 = \frac{1}{(N-2)}\sum_{n=1}^{N-2}(d_n - d_{n-1})^2 \tag{3}$$

These variances (3) are used to calculate the Hjorth mobility ($m_H$) and the Hjorth complexity ($c_H$) [29] as shown in (4).

$$m_H = \sigma_d/\sigma_a \text{ and } c_H = \sqrt{\left(\sigma_{dd}^2/\sigma_d^2\right) - \left(\sigma_d^2/\sigma_a^2\right)} \tag{4}$$

*Detrended fluctuation analysis (DFA)*

DFA has been introduced in identifying long range correlations in non-stationary time series data. By using a scaling exponent ($\alpha$), one can describe the significant autocorrelation properties of signals with a provision of capturing the non-stationary behaviour as well [30], [31]. The different values of $\alpha$ represents certain auto-correlation properties of the signal [30], [31]. For a value of less than 0.5, the signal is described as anti-correlated. A value of exactly 0.5 indicates uncorrelated (white noise) signal, whereas a value greater than 0.5 indicates positive autocorrelation in the signal. When $\alpha = 1$, the signal is indicated to be $1/f$ noise and a value of 1.5 indicates the signal to be random walk or Brownian noise [30], [31].

*Hurst exponent*

The Hurst exponent ($H$), a dimensionless estimator similar to DFA, is used as a measure of the long term memory of a time series data $x_i$ [32], [33]. The value of the Hurst exponent lies between 0 and 1, with a value between 0 – 0.5 indicating anti-persistent behaviour. This denotes that a decrease in the value of an element will be followed by an increase and vice versa. This characteristic is also known as mean reversion, which is explained as the tendency of future values to return to longer term mean value. The mean reversion phenomenon gets stronger for a series with exponent value closer to zero [32], [33]. When the value is close to 0.5, a random walk (e.g. a Brownian time series) is indicated. In such a time series, there is no correlation between any element and predictability of future elements is difficult [32], [33]. Lastly, when the value of the exponent is between 0.5 and 1, the time series exhibits persistent behaviour. This means the series has a trend or there is a significant autocorrelation in the signal. The more closer the exponent value gets towards unity, a stronger trend is indicated for the time series [32], [33].

*Wavelet entropy (Wentropy)*

The time series may be represented in frequency and/or time-frequency domains by decomposing the signal in terms of basis functions such as harmonic functions (as in Fourier analysis) or wavelet basis functions (with consideration of non-stationary behaviour), respectively. Given such decomposition, it is possible to consider the distribution of the expansion coefficients in this basis. Quantification of the degree of variability of the signal could be done using the entropy measure, where high values indicate less ordered distributions. The wavelet packet transform based entropy (WE) measures the degree of disorder (or order) in a signal [34–36]. A very ordered underlying process of a dynamical system may be visualized as a periodic single frequency signal (with a narrow band spectrum). Now the wavelet transformation of such a signal, will be resolved in one unique level with value nearing one, and all other relative wavelet energies being minimal (almost equal to zero) [34–36].

On the other hand, a disordered system represented by a random signal will portray significant wavelet energies from all frequency bands. The wavelet (Shannon) entropy gives an estimate of the measure of information of the probability distributions. This is calculated by converting the squared absolute values of the wavelet coefficients $s_i$ of the $i^{th}$ wavelet decomposition level as shown in (5).

$$WE = -\sum_i s_i^2 \log(s_i^2) \tag{5}$$

*Average spectral power*

The average spectral power ($\overline{P}$) is the measure of the variance of signal power, distributed across various frequencies [37]. It is given by the integral of the power spectral density (PSD) curve $\left|X(e^{j\omega})\right|^2$ of the signal $x(t)$ within a chosen frequency band of interest (bounded by the low and high frequency – $\omega_l$, $\omega_h$ respectively) as shown in (6).

$$\overline{P} = \int_{\omega_l}^{\omega_h} \left|X(e^{j\omega})\right|^2 d\omega \tag{6}$$

*(d) Adopted classification scheme*

All of the above mentioned extracted features are first normalized to scale them within a maximum (1) and minimum (0) value and to avoid any unnecessary emphasis of some of the features on the classifier weights due to their larger magnitude than the others. Amongst all the 11 features, their relative importance in each of the binary classification set has been obtained by computing the Fisher's Discriminant Ratio (FDR) [38]. The FDR is a measure to explore the discriminating power of a particular feature to separate two classes and are computed as $(\mu_1 - \mu_2)^2 / (\sigma_1^2 + \sigma_2^2)$ [38], where, $\mu_1$ and $\mu_2$ are the mean and $\sigma_1$ and $\sigma_2$ are the standard deviation of the features in the two classes respectively and therefore should not be confused with that of the raw signal in (1). Higher ranking, based on FDR, will be assigned to those features which have higher difference in the mean values and small standard deviation implying compact distantly located clusters. Due to the application of multiple-stimulus, the FDR based feature ranking is applied for each of the stimulus pairs, in the present work [38].

The classifiers implement algorithms which help in distinguishing between two or more different groups or classes of data. Different classification algorithms are obtained by first training the class labels (stimulus applied in this case) of a certain portion of the known (training) groups and then using the trained model to predict the class labels for a group of unknown (test) dataset. Once it is found that the testing phase is successful (high accuracy in identifying the stimulus) using the trained model, the algorithm can be used to identify which class an unknown data belongs to. In cases, where the distinction is easily achievable, discriminant analysis classifiers such as Linear Discriminant Analysis (LDA) could be effective. Where such distinctions are not that straightforward, nonlinear classifiers such as kernel based techniques like support vector machine (SVM) can be applied. Cases where only two groups need to be identified, binary classification are generally carried out. This is a much simpler process than multiple class classification. The choice of a classifier (discriminant or complex kernelized SVM) may be determined sometimes by looking at the distribution plot of the features of the two groups. If the distribution plots show two well separated means, we can conclude that a simple linear or other discriminant analysis based classifiers should be able to classify the data to a sufficient extent. Unnecessarily involving a complex nonlinear classification technique often gives high classification accuracy on the training dataset, but is prone to over-fitting. In the present study we focus on five different discriminant analysis classifiers which are based on least square method for training the classifier weights compared to the computationally heavy optimization process involved in SVM. Amongst five discriminant analysis variants the QDA uses a quadratic kernel with the feature vectors. The *Diaglinear* and *Diagquadratic* classifiers are also known as Naïve Bayes classifier using a simple linear and quadratic kernel and use the diagonal estimate of the covariance matrix (neglecting the cross-terms or feature correlations). The *Mahalanobis* classifier uses a different distance measure than the standard Euclidean distance [38]. We used different discriminant analysis classifiers due to their simplicity to see the characteristic changes traced in the features due to these stimuli. Two types of approach could be taken in classification – 1) choice of meaningful statistical features followed by simple classifier, 2) simple features followed by a complex classifier. The former case is preferable in the present case since it may help in understanding the change in statistical behaviour of the signal which might be indicative towards some consistent modification of the underlying biological process.

Cross validation schemes are often employed to avoid the introduction of any possible bias due to the training data-set [38]. Here we use the leave one out cross validation (LOOCV) where, if there are *N* data-points, then (*N-1*) number of samples is used for training the classifier and the one held-out sample is used to test the trained structure. Thereafter, the single test sample will be included in the next training set, and again a new sample from the previous training set will be set aside as the new test data. This loop will go for *N* times, till all the samples have been tested and the average classification accuracy for all the *N* instances are calculated [38].

### 3. Results

The classification results of 5 discriminant analysis classifier variants, using 11 statistical features from plant electrical signal response to four different stimuli *viz.* $H_2SO_4$, NaCl of 5 ml and 10 ml and $O_3$ are presented in this section. We also investigate which stimuli are best detected by looking at the classification accuracy, thereby suggesting the ability of the plant to detect few particular stimuli better than the others.

*a) Need for subtracting the background information of individual features*

Figure 3, shows the four plant electrical signal responses to four different stimuli beginning at different amplitude levels. This means the background signal (even before the

Journal of the Royal Society Interface

application of the stimuli) is different in all four different cases. This may bias the final classification result due to the already separated background information within the multiple features considered. Due to the effect of different backgrounds, we can see a clear separation between the stimuli for some features such as Hjorth mobility, Hjorth complexity and skewness, in Figure 4, where histogram plots for each of the features for each stimulus are plotted without any background subtraction.

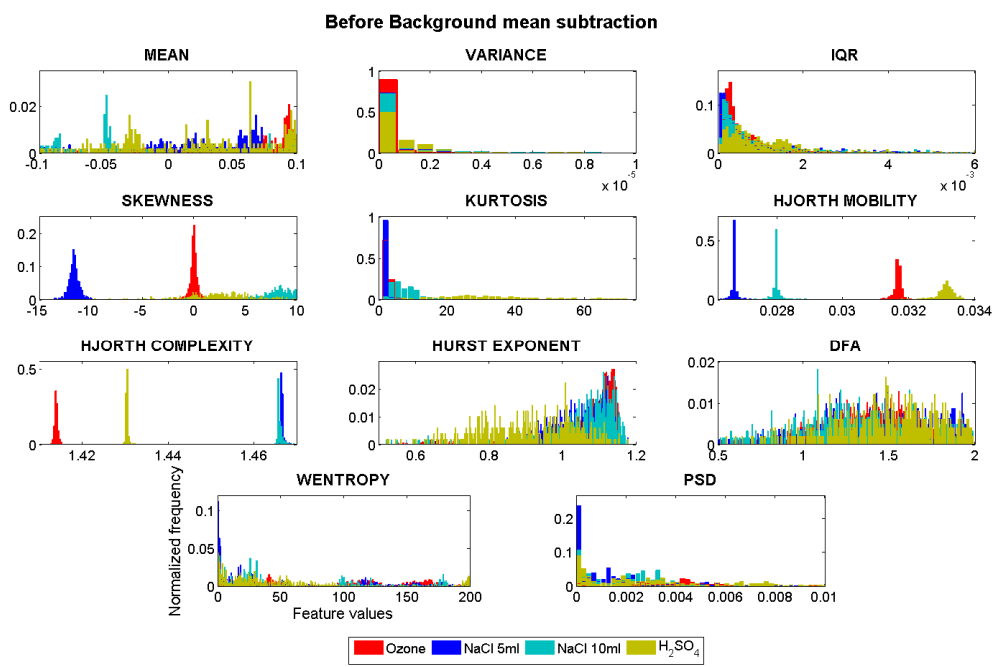

Figure 4: Normalized histogram plots for 11 individual features showing stimuli separability (no background subtraction).

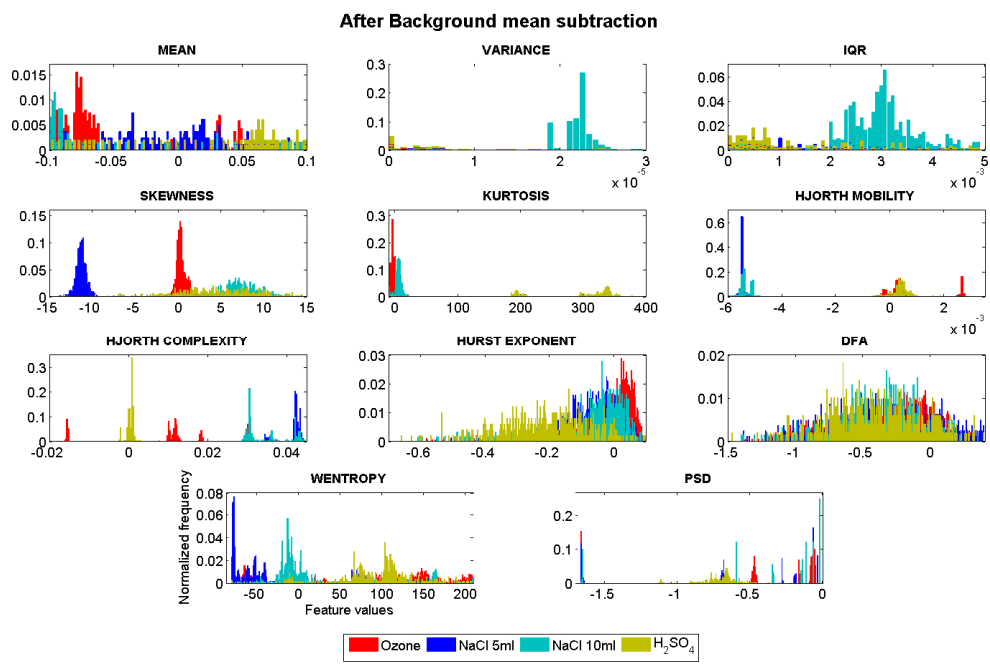

Figure 5: Univariate histograms of each of the 11 features for four different stimuli (with background subtraction)

This encourages us to look at only the *incremental values* of the features under different stimuli. The incremental values are obtained by subtracting the mean of every feature extracted from the background from the corresponding feature extracted from the post-stimulus part of the signal. The histogram plots of the incremental values of the individual features, after the background is subtracted, are given in Figure 5 which shows a lesser separability in the stimuli which were as expected. We now used these incremental values of the features to see how good they are in providing a successful classification (using five different discriminant analysis classifier variants) between any two stimuli (six binary combinations of four stimuli). As an example, although the histogram plots in Figure 5 shows clear separation of the distributions for NaCl and $O_3$ using skewness as feature due to their peaky nature, the frequency of occurrence of the histograms show that the distributions have wider spread which has been reflected by the moderate rate of classification reported in the next subsections using that particular feature.

b) *Correlation of features to avoid redundancy*

Table 2: Correlation coefficient between 11 statistical features extracted from plant electrical signals (after subtracting the mean of the pre-stimulus features from the post-stimulus ones)

| Features | $\mu$ | $\sigma^2$ | IQR | $\gamma$ | $\beta$ | $m_H$ | $c_H$ | $H$ | $\alpha$ | WE | $\bar{P}$ |
|---|---|---|---|---|---|---|---|---|---|---|---|
| $f_1 = \mu$ | 1.00 | 0.09 | -0.03 | -0.06 | 0.07 | 0.04 | 0.03 | -0.11 | -0.22 | 0.70 | 0.26 |
| $f_2 = \sigma^2$ | * | 1.00 | 0.83 | 0.01 | 0.10 | -0.05 | -0.23 | -0.10 | 0.21 | 0.02 | 0.07 |
| $f_3 = IQR$ | * | * | 1.00 | -0.04 | 0.02 | -0.08 | -0.31 | 0.01 | 0.53 | -0.12 | 0.05 |
| $f_4 = \gamma$ | * | * | * | 1.00 | 0.29 | 0.00 | -0.06 | -0.09 | -0.07 | -0.08 | 0.00 |
| $f_5 = \beta$ | * | * | * | * | 1.00 | -0.01 | -0.23 | -0.14 | -0.03 | 0.14 | 0.06 |
| $f_6 = m_H$ | * | * | * | * | * | 1.00 | 0.34 | -0.07 | -0.10 | 0.06 | 0.02 |
| $f_7 = c_H$ | * | * | * | * | * | * | 1.00 | -0.12 | -0.28 | 0.06 | 0.09 |
| $f_8 = H$ | * | * | * | * | * | * | * | 1.00 | 0.64 | -0.15 | -0.16 |
| $f_9 = \alpha$ | * | * | * | * | * | * | * | * | 1.00 | -0.29 | -0.06 |
| $f_{10} = WE$ | * | * | * | * | * | * | * | * | * | 1.00 | -0.09 |
| $f_{11} = \bar{P}$ | * | * | * | * | * | * | * | * | * | * | 1.00 |

Between all the features, a correlation test was carried out to find out their inter-dependence. The result of this test, given in Table 2, is obtained by checking the Pearson correlation coefficient values between all feature pairs. A correlation value of (~ +1/-1) indicates a strong positive/negative correlation between a pair, whereas a value closer to zero indicates the feature pairs are independent and are thus more informative about the underlying process. A good classification strategy should ideally involve uncorrelated features, in order to avoid redundancy in training the classifier. In this work, we proceeded by initially taking all features into account and then ignored the ones with high correlation.

c) *Classification using univariate and bivariate features*

The classification results were obtained in two ways – using univariate and bivariate features, to make the analysis intuitive and simple to infer. That is, instead of taking all the features together to get a multivariate classification (which may give good classification accuracy but are less intuitive and reliable due to increase in complexity and dimension of the problem), we just explored the results with 11 individual features and 55 possible feature pairs.

Table 3 presents the results, obtained using individual features, averaged across all the six stimuli combinations and all the five different classifier variants. We have also presented the relative multi-class separability score given by the scatter matrix ($S$) in (7) for each feature in terms of the within-class ($S_w$) and between-class ($S_b$) scatter matrix [38].

$$S = trace\left(S_w^{-1} S_b\right)$$
$$S_w = \sum_{i=1}^{c} P_i S_i, S_b = \sum_{i=1}^{c} P_i (\mu_i - \mu_0)(\mu_i - \mu_0)^T, \mu_0 = \sum_{i=1}^{c} P_i \mu_i, i = 1, 2, \cdots, c \quad (7)$$

Here, $P_i$ is the a-priori probability for the present four class problem ($c=4$) and has been considered as ¼. Also, the mean and covariance matrices for each of the classes are denoted by $\{\mu_i, S_i\}$ and the $\mu_0$ is the global mean vector. The scatter matrix extends the concept of class-separability using FDR from binary classification to multi-class problems.

Table 3: Average accuracy (averaged across all six binary stimuli combinations and all five classifier variants) and best accuracy (averaged across four *one vs. rest* stimuli combinations) for classification using individual features

|  | Ranked Features | Scatter matrix | Average accuracy for all *binary* stimulus combinations (%) | Best accuracy for all *'one vs. rest'* stimulus combinations (%) |
|---|---|---|---|---|
| $F_1$ | Mean ($\mu$) | 0.8453 | 70.87 | 73.01, *Mahalanobis* |
| $F_2$ | Wentropy ($WE$) | 0.2858 | 69.79 | 62.26, *Mahalanobis* |
| $F_3$ | Hjorth Complexity ($c_H$) | 0.1022 | 66.61 | 60.82, *Mahalanobis* |
| $F_4$ | Inter Quartile Range ($IQR$) | 0.2838 | 65.07 | 63.62, *LDA/Diaglinear* |
| $F_5$ | Variance ($\sigma^2$) | 0.0453 | 63.57 | 65.58, *LDA/Diaglinear* |
| $F_6$ | Average spectral power ($\bar{P}$) | 0.1385 | 60.51 | 61.58, *Mahalanobis* |
| $F_7$ | DFA ($\alpha$) | 0.1989 | 60.14 | 61.28, *Mahalanobis* |
| $F_8$ | Kurtosis ($\beta$) | -0.0637 | 58.06 | 62.64, *Mahalanobis* |
| $F_9$ | Hjorth Mobility ($m_H$) | 0.0064 | 57.45 | 61.44, *Mahalanobis* |
| $F_{10}$ | Skewness ($\gamma$) | -0.5731 | 54.55 | 62.09, *Mahalanobis* |
| $F_{11}$ | Hurst Exponent ($H$) | 0.0321 | 52.38 | 61.40, *Mahalanobis* |

The scatter matrices value in Table 3 provide an insight about how good the separation between all the four classes (stimuli) are using the individual features. From Table 3 we can see that the signal 'mean' on its own has the best classification result for all the six binary combinations of four stimuli. However since we have extracted the features from the raw non-stationary plant electrical signals, mean is not a very reliable feature to base any conclusions on, because it can be influenced by various artefacts and noise during measurement or from various environmental factors (e.g. sudden gust of breeze could shake the electrodes connected to the plant body etc.). The next five best features (best average accuracies given in Table 3), when taken individually, are wavelet packet entropy, Hjorth complexity, interquartile range, variance and average spectral power respectively. From now on, we will only consider these features as the top five features. In Table 3, we also report the best achievable accuracy along with the best classifier using each of the single features to discriminate the four stimulus within a '*one vs. rest*' strategy. This highlights the possibility of isolating one particular class from the other classes using a single feature, with a certain degree of confidence.

So far we have seen the averaged results of classification for six binary stimuli combinations using individual features. We next find the best classified stimuli combination using only the top five individual features and using the five variants of the discriminant analysis classifiers, as mentioned above. As a result, we obtained five classification accuracies (for five individual features) for every classifier for each of the six binary stimuli combination. That results in 25 classification accuracies for each of the six binary stimuli combinations. All these 25 results were averaged for each stimuli combination and given in Table 4 which shows the best discrimination possible is for $H_2SO_4$ and $O_3$ with classification accuracy over 73%. Additionally, discrimination between NaCl (both concentrations) and $O_3/H_2SO_4$ also shows promising result with accuracy over 65% and 63% respectively.

Table 4: Accuracy using top five individual (univariate) features ($F_2$ through $F_6$) and averaged across five classifiers (average separability between different stimulus combinations)

| Stimuli | NaCl 5ml | NaCl 10ml | $H_2SO_4$ | $O_3$ |
|---|---|---|---|---|
| NaCl 5ml | - | 57.20% | 64.02% | 65.94% |
| NaCl 10ml | * | - | 63.17% | 67.29% |
| $H_2SO_4$ | * | * | - | 73.03% |
| $O_3$ | * | * | * | - |

The average classification results presented in Table 4 encourages us to look at the best results achieved using individual features, for each stimuli combination, so that we can see if there is any consistent feature giving good classification results. This is shown in Table 5 from where it is evident that $F_3$ (Hjorth complexity) gives the best result for three different binary stimuli combinations with an accuracy over 74%. Overall, the best accuracy is achieved for classification between $H_2SO_4$ and $O_3$, with an accuracy of >94% using $F_2$ (Wentropy) and QDA classifier. Although in Table 4 the discrimination between NaCl and $O_3/H_2SO_4$ are shown in terms of the average accuracy which might seem to be relatively low (63% or 65%), the best cases for such a discrimination can be found in Table 5 (accuracies of >78% and >72% respectively) between the same set of stimuli. Also from Figure 5 we can see that though skewness shows a good discrimination between $H_2SO_4$ and other stimuli, from Table 3 we can see that the average classification accuracy using skewness as an individual feature is very low. This is due to the fact that skewness on its own did not give good classification results between other remaining stimuli combinations.

Table 5: Best accuracy taking individual features for each stimulus combinations (best separability between different stimulus combinations)

| Stimuli | NaCl 5ml | NaCl 10ml | $H_2SO_4$ | $O_3$ |
|---|---|---|---|---|
| NaCl 5ml | - | 74.36% ($F_3$, LDA classifier) | 75.09% ($F_3$, Mahalonobis classifier) | 78.95% ($F_3$, LDA classifier) |
| NaCl 10ml | * | - | 72.13% ($F_8$, LDA classifier) | 82.27% ($F_9$, QDA classifier) |
| $H_2SO_4$ | * | * | - | 94.95% ($F_2$, QDA classifier) |
| $O_3$ | * | * | * | - |

### d) Classification using feature pairs

Table 6: Average accuracy obtained using top five feature pairs (bivariate) and five classifiers (average separability between different stimulus combinations)

| Stimuli | NaCl 5ml | NaCl 10ml | $H_2SO_4$ | $O_3$ |
|---|---|---|---|---|
| NaCl 5ml | - | 59.52% | 58.21% | 72.69% |
| NaCl 10ml | * | - | 64.66% | 76.60% |
| $H_2SO_4$ | * | * | - | 74.60% |
| $O_3$ | * | * | * | - |

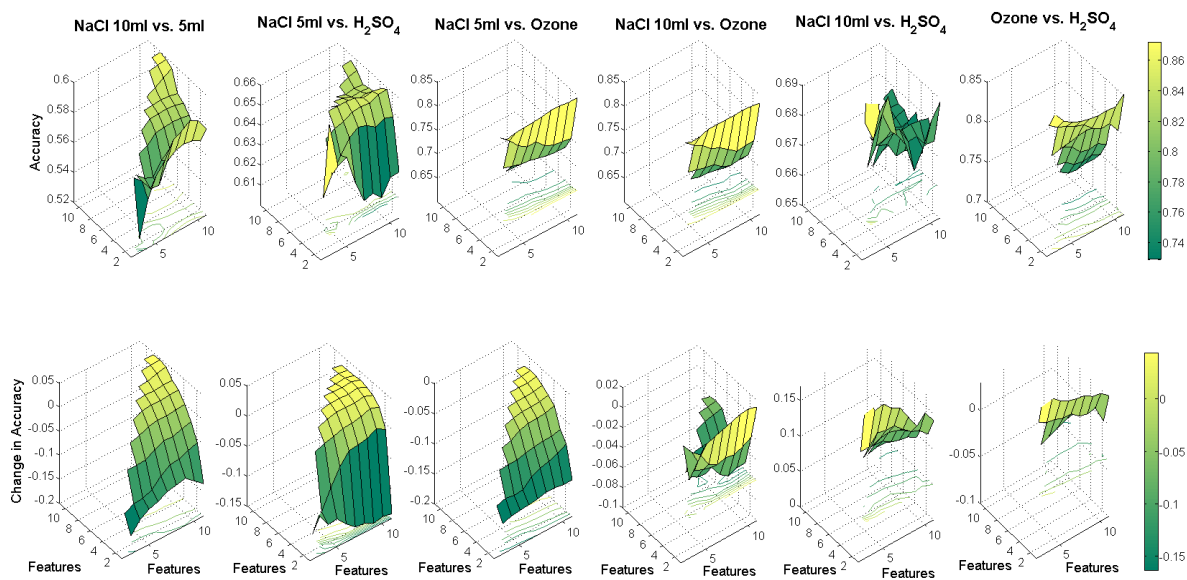

Figure 6: (top) Classification accuracy for different feature combinations with background information removed; (bottom) deterioration in accuracy for the features without background information removed.

Next, we looked at the effect of all possible feature pairs using 11 individual features (totalling 55 independent feature pairs) on the classification results between six different stimuli combinations. These classification accuracies are shown in Figure 6 along with the difference in accuracy (error) when the background is not subtracted as discussed in earlier section. The features mentioned as $\{1, 2, \cdots, 11\}$ in Figure 6 are the features designated by $\{f_1, f_2, \cdots, f_{11}\}$ respectively in Table 2. Since we ignored mean as a feature in the previous section, we explored the effect of taking binary combinations of the next five individual features ($F_2$ through $F_6$, as mentioned in Table 3) on the classification accuracy. The results obtained using each of these bivariate features (pairs), using all the five classifier variants were averaged and given in Table 6 which are found to be better than the averaged results obtained using just univariate features as given in Table 4.

Table 7: Best accuracy for each stimulus combination using two features (best separability between different stimulus combinations)

| Stimuli | NaCl 5ml | NaCl 10ml | $H_2SO_4$ | $O_3$ |
|---|---|---|---|---|
| NaCl 5ml | - | 63.18% ($F_4$-$F_6$ with Diaglinear) | 65.87% ($F_4$-$F_6$ with linear) | 82.69% ($F_4$-$F_5$ with Diaglinear) |
| NaCl 10ml | * | - | 73.18% ($F_4$-$F_5$ with Diagquadratic) | 92.06% ($F_4$-$F_5$ with Mahalanobis) |
| $H_2SO_4$ | * | * | - | 87.48% ($F_4$-$F_5$ with Quadratic) |
| $O_3$ | * | * | * | - |

By this exploration, we wanted to find out if there is any improvement on the classification accuracy when a feature pair is used rather than just individual feature. From Table 6, we can see that classification accuracy is improved for all stimuli combination except NaCl 5ml vs. $H_2SO_4$. We can also observe from Table 6 that the top two best accuracies are obtained for stimuli combinations of NaCl 10ml vs. $O_3$ and $H_2SO_4$ vs. $O_3$. Now, let us look at the best feature pair(s), among all 55 bivariate feature pairs, as given in Table 7. We notice that a combination of $F_4$ (IQR) and $F_5$ (variance) results in the best classification accuracies for four out of six different stimuli combinations. For the remaining two stimuli combinations, a feature pair of $F_4$ and $F_6$ (average spectral power) gives the best classification accuracies.

*e) Finding the most reliable combination of feature or feature pair and classifier variant*

So far we have found that individual features $F_2, F_3, F_8$ and $F_9$ and feature pairs $F_4$-$F_5$ and $F_4$-$F_6$ produced the best classification results for one or more (out of the six) stimuli combinations. We now explore these features and feature pairs for all stimuli combinations. Table 8 gives the results of classification when we used just $F_2, F_3, F_8$ and $F_9$ as an individual feature using all classifier variants for all binary stimuli combinations. Similarly, Table 9 gives the results using the feature pairs $F_4$-$F_5$ and $F_4$-$F_6$ for all the six stimuli combinations, using all the five classifier variants. These results will help us choosing the right classifier and deciding the feature or feature-pair which provides the best average accuracy for all the binary combinations of stimuli.

Table 8: Accuracy of different classifiers for six stimuli combinations (in %) using the best individual features

| Individual feature | Classifier variant | NaCl 5ml vs 10ml | NaCl 5ml vs $H_2SO_4$ | NaCl 5ml vs $O_3$ | NaCl 10ml vs $O_3$ | NaCl 10ml vs $H_2SO_4$ | $O_3$ vs $H_2SO_4$ | Average Accuracy (%) |
|---|---|---|---|---|---|---|---|---|
| $F_2$ (Wentropy) | LDA | 55.3 | 66.4 | 73.4 | 77.6 | 59.5 | 82.8 | 69.2 |
| | QDA | 52.2 | 67.6 | 62 | 74 | 56.4 | 95 | 67.9 |
| | Diaglinear | 55.3 | 66.4 | 73.4 | 77.6 | 59.5 | 82.8 | 69.2 |
| | Diagquadratic | 52.2 | 67.6 | 62 | 74 | 67.3 | 95 | 69.7 |
| | Mahalanobis | 55.5 | 73.1 | 73.7 | 78.2 | 63.8 | 94.4 | *73.1* |
| $F_3$ (Hjorth Complexity) | LDA | 74.4 | 73.9 | 78.9 | 66 | 68.3 | 66.9 | 71.4 |
| | QDA | 74.1 | 47.1 | 61.6 | 71.5 | 67.5 | 41.8 | 60.6 |
| | Diaglinear | 74.4 | 73.9 | 78.9 | 66 | 68.3 | 66.9 | 71.4 |
| | Diagquadratic | 74.1 | 47.1 | 61.6 | 71.5 | 67.5 | 41.8 | 60.6 |
| | Mahalanobis | 74.4 | 75.1 | 62.4 | 51.1 | 69.4 | 81.5 | 69 |
| $F_8$ (Kurtosis) | LDA | 57.1 | 66.5 | 57.6 | 60.5 | 72.1 | 66.1 | 63.3 |
| | QDA | 47.8 | 38.4 | 69.3 | 47.1 | 71.8 | 22.5 | 49.5 |
| | Diaglinear | 57.1 | 66.5 | 57.6 | 60.5 | 72.1 | 66.1 | 63.3 |
| | Diagquadratic | 47.8 | 38.4 | 69.3 | 47.1 | 71.8 | 22.5 | 49.5 |
| | Mahalanobis | 57.7 | 58.2 | 38.5 | 81.3 | 71.3 | 81 | 64.7 |
| $F_9$ (Hjorth Mobility) | LDA | 60.4 | 73.9 | 68.5 | 66 | 56 | 35.8 | 60.1 |
| | QDA | 49.7 | 47.7 | 76.4 | 82.3 | 48.7 | 81.6 | 64.4 |
| | Diaglinear | 60.4 | 48.8 | 68.5 | 66 | 56 | 35.8 | 55.9 |
| | Diagquadratic | 49.7 | 47.7 | 76.4 | 82.3 | 48.7 | 81.6 | 64.4 |
| | Mahalanobis | 66.2 | 54.6 | 30.1 | 24.7 | 58.4 | 20.5 | 42.4 |

From Table 8, we note that using just $F_2$ or $F_3$ provides consistently better average classification accuracies than using $F_8$ or $F_9$. It is also noticed that although $F_2$ provides a better classification for the stimuli combinations NaCl 10ml vs. $O_3$ and $O_3$ vs. $H_2SO_4$, $F_3$ provides much consistent and better result for the remaining stimuli combinations. While considering a single feature for discriminating the four stimuli, the best average result (73%) could be obtained using the $F_2$ (Wentropy) as feature and Mahalanobis classifier, although it is highly correlated with the signal mean ($F_1$) as shown in Table 2. Since mean as a feature was ignored due to its susceptibility to artefacts, therefore, we also ignore Wentropy and instead propose Hjorth complexity as the best individual feature for achieving good average classification accuracy.

ng the individual features. From ‖ REF _Ref386557847 m Table 9 that the top two classification accuracies (>73%) are obtained using $F_4$-$F_5$ combination and Diagquadratic and QDA as classifiers respectively. Both these top bivariate classification results (average accuracy of 69.65% across all stimuli and classifiers) are better than that obtained in univariate case in Table 8 (average accuracy of 62.98% across all stimuli and classifiers). Although again from Table 2, we realize the IQR and variance are highly correlated with each other but since we are achieving a good result in terms of classification using these two features, we note that calculating IQR and variance from a block of 1000 samples of raw non-stationary plant electrical signal, along with QDA or Diagquadratic classifier will provide consistently good results in terms of classifying which external stimuli caused the particular signature in the plant electrical signal.

Table 9: Accuracy of different classifiers for six stimuli combinations (in %) using the best feature pairs

| Best feature set | Classifiers | NaCl 5ml vs 10ml | NaCl 5ml vs $H_2SO_4$ | NaCl 5ml vs $O_3$ | NaCl 10ml vs $O_3$ | NaCl 10ml vs $H_2SO_4$ | $O_3$ vs $H_2SO_4$ | Average Accuracy (%) |
|---|---|---|---|---|---|---|---|---|
| $F_4 - F_5$ (IQR-Variance) | LDA | 56.61 | 62.61 | 82.53 | 81.57 | 69.24 | 85.09 | 72.94 |
| | QDA | 57.78 | 61.62 | 82.24 | 86.06 | 64.69 | 87.49 | 73.31 |
| | Diaglinear | 63.17 | 53.8 | 82.7 | 83.47 | 62.06 | 80.33 | 70.92 |
| | Diagquadratic | 60.34 | 58.43 | 82 | 86.08 | 73.19 | 81.98 | *73.67* |
| | Mahalanobis | 62.17 | 50.23 | 80.24 | 92.07 | 53.64 | 78.91 | 69.54 |
| $F_4 - F_6$ (IQR-Average spectral power) | LDA | 56.88 | 65.87 | 71.61 | 70.06 | 67.94 | 81.27 | 68.94 |
| | QDA | 57.82 | 57.01 | 73.89 | 81 | 65.19 | 78.48 | 68.9 |
| | Diaglinear | 63.19 | 54.59 | 68.67 | 78.16 | 64.92 | 71.39 | 66.82 |
| | Diagquadratic | 58.11 | 60.57 | 79.19 | 75.70 | 67.96 | 79.66 | 68.31 |
| | Mahalanobis | 62.61 | 52.99 | 62.72 | 79.62 | 58.08 | 60.84 | 63.20 |

We next explore some pairs of uncorrelated features for classification by looking at the 12 next best average classification accuracies (obtained across all stimuli combinations and using all five different classifiers) as shown through a 2D normalized histogram (volume being unity) plots showing the separation of the four stimuli in Figure 7. Average accuracy obtained (over all stimuli combinations and classifiers) using particular feature pairs (denoted by $f_1, f_2, \cdots, f_{11}$, as described in Table 2) are also mentioned in the title of each subplot in Figure 7. It is observed that the second best average classification accuracy is achieved using variance and skewness as features which are almost uncorrelated (correlation of ~0.01 in Table 2).

In Figure 7, except the first subplot with $f_2 - f_3$, all the rest combinations are almost uncorrelated and still give a good classification performance. Thus as a reliable measure of analysis, it has been found that the *variance* and *skewness* calculated from a block of 1000 samples of plant electrical signal will be able to give an average (over all six stimuli combinations and using all five discriminant classifiers) accuracy of 70% during binary classification of the stimuli. It is to be noted that in the bivariate classification scheme, the mean ($F_1$) has not been considered as one of the features. Also, the best bivariate accuracies were achieved involving the variance ($F_2$) along with all the other features ($F_4...F_{11}$) in Figure 7, while ignoring the $F_2$-$F_3$ combination due to their high inter-dependence. As a summary, a better reliable classification scheme is expected (~67%-70%) involving bivariate features as shown in Figure 7, with respect to the univariate features as given in Table 3 (<67%, ignoring mean and Wentropy).

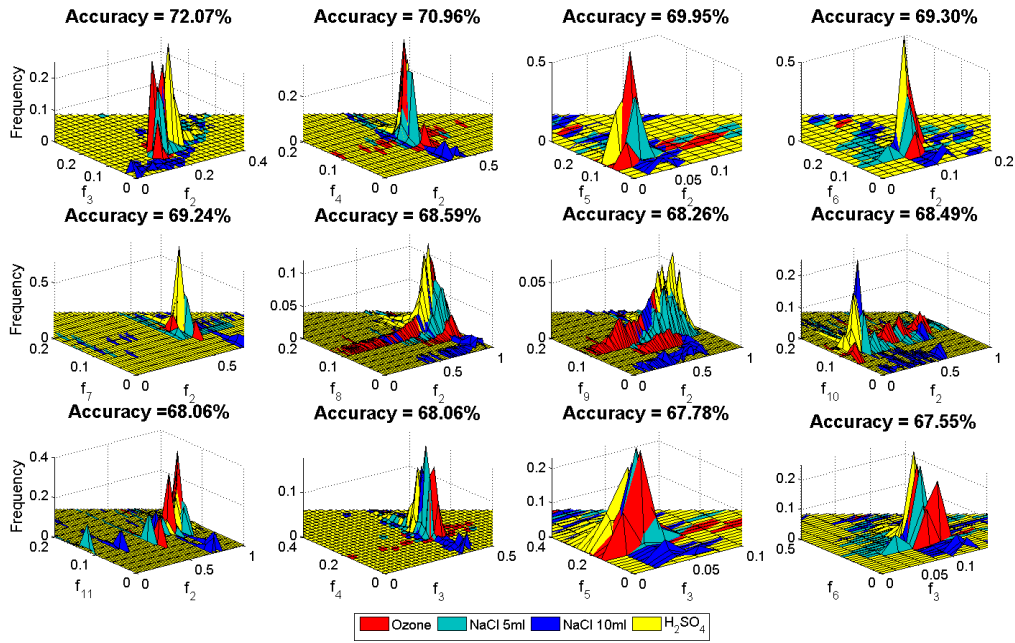

Figure 7: Bivariate histograms of top feature pairs with highest classification accuracy for all the four stimuli (accuracy mentioned in title of each subplot).

The results presented in this work only takes into account the experimental data for individual stimulus under controlled environment (laboratory). The next step can be to setup experiments where multiple stimuli could be applied together on the plant and its electrical signal response could be extracted for further analysis and classification of the most influential stimuli. Also, the robustness of the statistical features due to possible artefact (e.g. movement of the leaf due to wind, rainfall etc.) are to be explored in future in a more naturalistic environment, outside the controlled laboratory set-up.

## 4. Discussion

In our exploration, the data from two channels per plant (per experiment) were used to record the electrical response, and then statistical features were calculated from both the channels and pooled together. Here, the location on the plant body for the data extraction was ignored, as the work was primarily focussed more on the possibility of classification of applied external stimuli from the extracted plant electrical signal. Similarly the effect of a different species of plants to study all the four stimuli have also been ignored, except the introduction of an additional species (cucumber) for Ozone stimulus. The idea behind developing an external stimuli classification scheme, based on plant electrical response is focussed on generic plant signal behaviour and not of a specific species. However such isolation forms a very good study and could be taken up as future scope of work. There might be some possibility of confounding effects based on the position of the electrodes and plant species in any classification scheme. But such confounding effects will be minimal due to the large number of data samples as shown in Table 1 and the use of cross-validation scheme to test the performance of the discriminant analysis classifiers. Also we did not use kernel based nonlinear classifiers like SVM which could over-fit these plant specific characteristics and still give good classification result, rendering the loss of generalizing capability of the classifier.

Moreover, the present classification scheme is based on the raw non-stationary plant signal. In bio-signal processing literature [11], using of high-pass filter is recommended to make a bio-signal stationary instead of extracting features from the raw non-stationary signal. But there is also a possibility with an ad-hoc filtering that some useful information in the data may get lost since the cut-off frequency for plant signal processing is not yet known. That is why we considered the features from the post-stimulus signal to train the classifier by removing any possible bias of the channel or plant using incremental features i.e. using the mean of the features in the pre-stimulus part. The segmentation of the signal in a block of 1000 samples also disregards the temporal information of the stimuli, since we primarily tried to answer the question if classification is indeed possible by looking at any segment of the post-stimulus part of the signal. Also, in a realistic scenario, we would not know when the response to a particular stimulus started. So we need to base our classification on the in-coming stream of live data.

## 5. Conclusion

Our exploration using raw electrical signals from plants provides a platform for realizing a plant signal based bio-sensor to classify the environmental stimuli. The classification scheme was based on 11 statistical features extracted from segmented plant electrical signals, followed by feature ranking and rigorous univariate and bivariate feature based classification using five different discriminant analysis classifiers. External stimuli like $H_2SO_4$, $O_3$ and NaCl in two different amounts (5 ml and 10 ml) have been classified using the adopted machine learning approach with 11 statistical features, capturing both the stationary and non-stationary behaviour of the signal. The classification has yielded a best average accuracy of 70% (across all stimuli and five classifier variants using variance and skewness as feature pairs) and the best individual accuracy of 73.67% (across all stimuli and using variance and IQR as feature pairs in Diagquadratic classifier). The very fact that, by looking at the statistical features of plant electrical response, we can successfully detect which stimuli caused the signal is quite promising. This will not only open the possibility of remotely monitoring the environment of a large geographical area, but will also help in taking timely preventive measures for natural or man-made disasters.

**Acknowledgements**

The work reported in this paper was supported by project PLants Employed As SEnsor Devices (PLEASED), EC grant agreement number 296582, URL: http://pleased-fp7.eu/.

**Data accessibility**

The experimental data is available in the PLEASED website at http://pleased-fp7.eu/?page_id=253.